\documentstyle[emulate_apj,apjfonts,epsf]{article}
\gdef\h50min{$h_{50}^{-1}$}

\gdef\1054{MS\,1054$-$03}
\gdef\2053{MS\,2053$-$04}
\gdef\hmpc3{h$^3$ Mpc$^{-3}$}
\gdef\omit#1{}
\gdef\Labbe{Labb\'{e}}
\lefthead{Franx et al.}
\righthead{A Significant Population of Red, Near-IR selected High Redshift Galaxies}

\begin{document}

\submitted{Accepted for publication in ApJ Letters}
\title{A Significant Population of Red, Near-IR selected High Redshift Galaxies
\altaffilmark{1}}
\author{
Marijn Franx\altaffilmark{2},
Ivo Labb\'{e}\altaffilmark{2}, 
Gregory Rudnick\altaffilmark{3}, 
Pieter G. van Dokkum\altaffilmark{4}, 
Emanuele Daddi\altaffilmark{5}
Natascha M. F\"{o}rster Schreiber\altaffilmark{2}, 
Alan Moorwood\altaffilmark{5}, 
Hans-Walter Rix\altaffilmark{6}, 
Huub R\"{o}ttgering\altaffilmark{2}, 
Arjen van de Wel\altaffilmark{2}, 
Paul van der Werf\altaffilmark{2}, 
Lottie van Starkenburg\altaffilmark{2}, 
}

\altaffiltext{1}{Based on service mode observations collected at 
the European Southern Observatory, Paranal, Chile 
(ESO Programme 164.O-0612). 
}

\altaffiltext{2}{Leiden Observatory, P.O. Box 9513, NL-2300 RA,
Leiden, The Netherlands}

\altaffiltext{3}{Max-Plank-Institut f\"ur Astrophysik, P.O. Box 1317,
D-85741, Garching, Germany}

\altaffiltext{4}{California Institute of Technology, MS 105-24,
Pasadena CA 91125, USA}

\altaffiltext{5}{European Southern Observatory, D-85748, Garching,
Germany }

\altaffiltext{6}{Max-Plank-Institut f\"ur Astronomie, D-69117,
Heidelberg, Germany }

\begin{abstract}

We use very deep near-infrared photometry of the Hubble Deep
Field South taken with ISAAC/VLT to identify a population of high 
redshift galaxies with
rest-frame optical colors similar to those of  nearby
galaxies.
The galaxies are chosen by their infrared colors $J_s-K_s > 2.3$,
aimed at selecting galaxies with redshifts above 2.
When applied to our dataset, we find  14 galaxies with $K_s  < 22.5$,
corresponding to a surface density of $3 \pm 0.8$  arcmin$^{-2}$.
The photometric redshifts all lie above 1.9, with a median of 2.6 and
a rms of 0.7. 
The spectral energy distributions of these
galaxies show a wide range: one is very blue in the rest-frame UV,
and satisfies the normal Lyman-break criteria for high redshift,
star-forming galaxies. Others are quite red throughout the observed
spectral range, and are extremely faint in the optical, with a median
$ V = 26.6$.
Hence these galaxies would not be included in photometric samples
based on optical ground-based data, and spectroscopic follow-up is 
difficult.
The spectral energy distributions often show a prominent break, 
identified as the Balmer break or 4000 \AA\ break. The median age
is 1 Gyr when fit with a constant star formation model with dust, or 0.7 Gyr
when fit with a single burst model. Although significantly younger
ages cannot be excluded when a larger range of models is allowed, the
results indicate
that these galaxies are among the oldest 
at these redshifts. 
The volume  density to $K_s=22.5$ is  half that
of Lyman-break galaxies at $z \approx 3$. Since the
mass-to-light ratios of the red galaxies are likely to be higher, the 
stellar mass density
is inferred to be  comparable to that of Lyman-break galaxies.
These red galaxies may be the descendants of  galaxies which
started to form stars at very high redshifts, and they may evolve
into the most massive galaxies at low redshift.

\end{abstract}


\section{Introduction}

\def\figa{
\vbox{
\begin{center}
\leavevmode
\hbox{%
\epsfxsize=8truecm
\epsffile{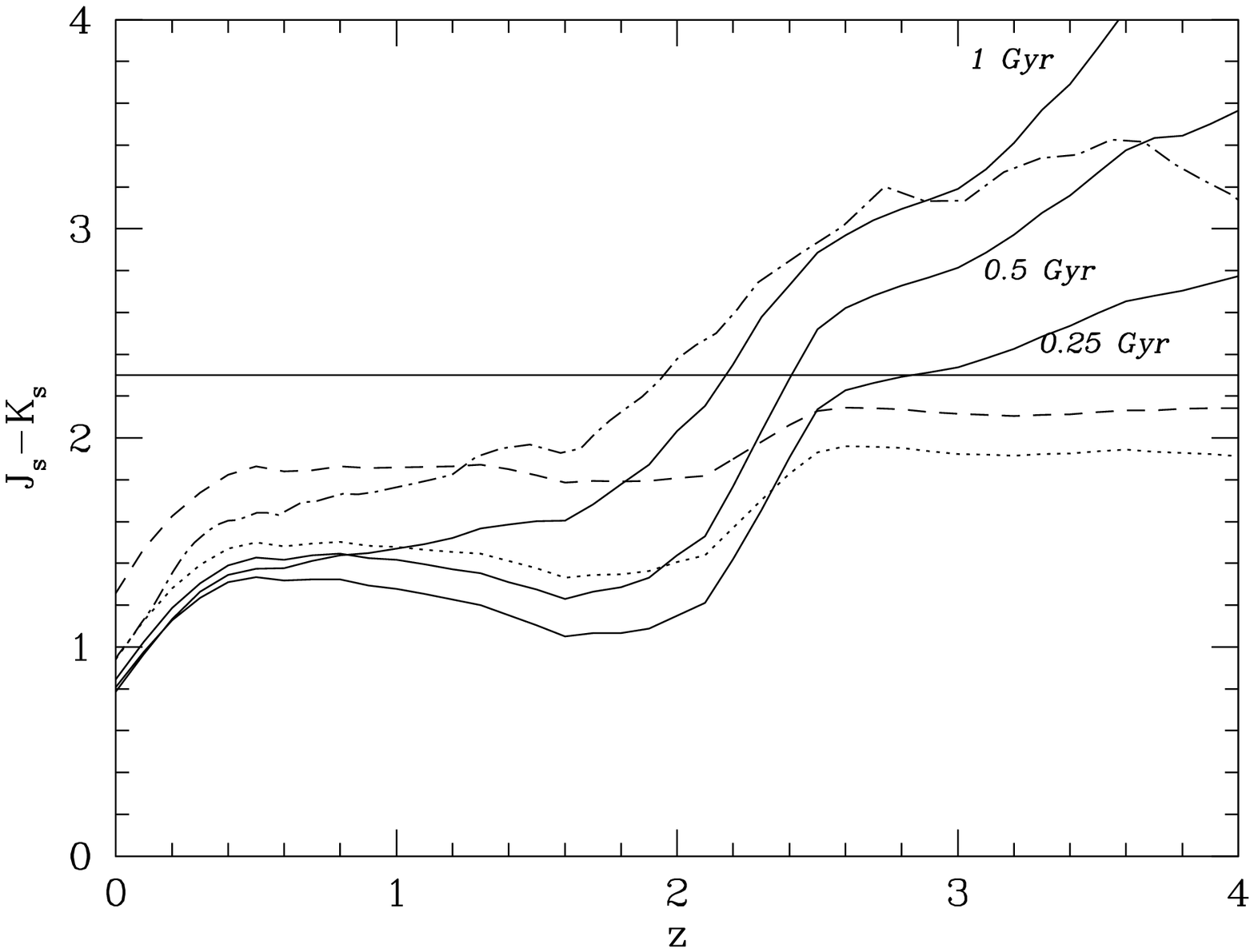}}
\figcaption{\small %
The $J_s-K_s$ color as a function of redshift for several galaxy
spectra.
The continuous curves indicate single age stellar populations with ages 
of 0.25, 0.5 and 1 Gyr. 
The colors exceed $J_s-K_s = 2.3$ only for $ z > 2$, due to the
Balmer break/4000 \AA\ break moving into the $J_s$ band.
The dotted and dashed lines indicate  models with continuous starformation
with ages and reddening of 1Gyr, E(B-V)=0.15, and 100Myr, E(B-V)=0.5, 
respectively.
Many galaxies with continuous star formation will not reach  $J_s - K_s
= 2.3 $, unless they are even older, or have larger reddening.
The dash-dot curve indicates the color evolution of a single burst
population which formed at $z=5$, and it also satisfies the color
criterion above $z=2$.}
\end{center}}}

\def\figb{
\vbox{
\begin{center}
\leavevmode
\hbox{%
\epsfxsize=8truecm
\epsffile{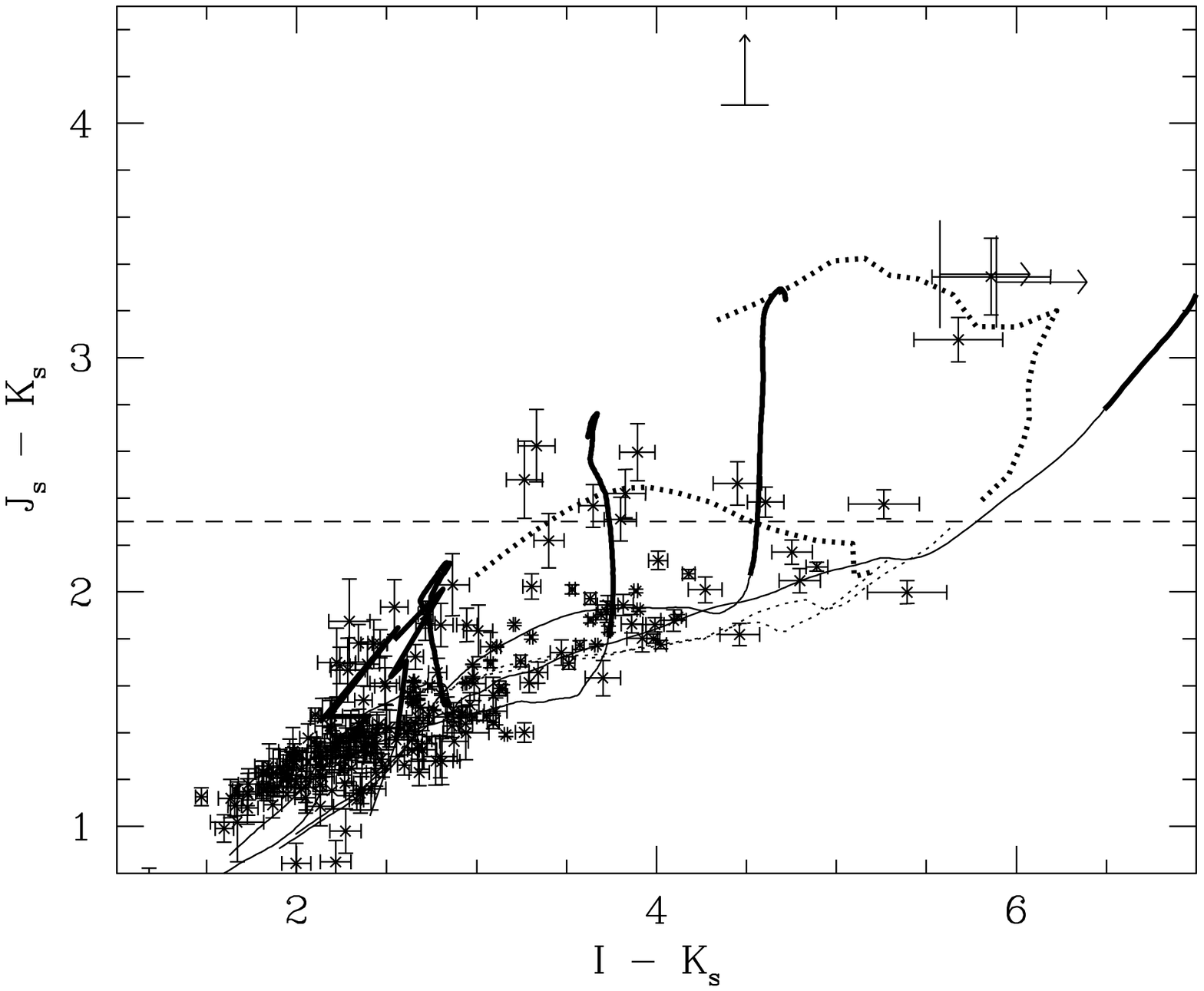}
}
\figcaption{\small
The $J_s-K_s$ colors versus $I-K_s$ colors of $K$-band selected
galaxies in the HDF-South.  A significant population of galaxies with
$J_s - K_s > 2.3$ is present, with a large range in $I - K_s$ colors.
The continuous curves indicate color tracks obtained by redshifting the
galaxy spectra measured by Coleman et al. (1980).  The dotted curves
indicates single age stellar populations with formation redshifts of
5, and 3. The curves are drawn thick for $2 < z < 4$.
}
\end{center}}
}

\def\figc{
\vbox{
\begin{center}
\leavevmode
\hbox{%
\epsfxsize=8truecm
\epsffile{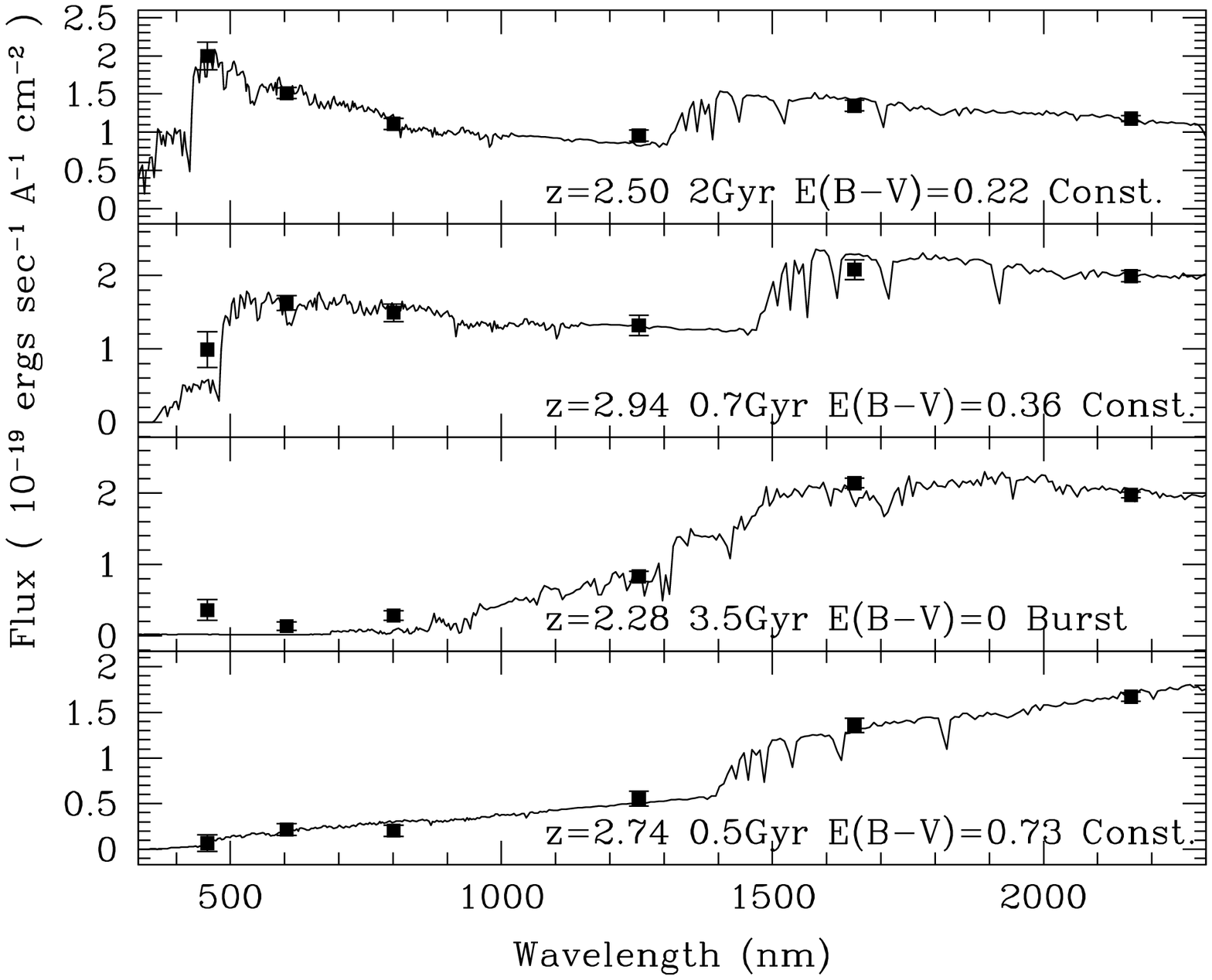}
}
\figcaption{\small
The spectral energy distributions  4 of the galaxies with $J_s-K_s > 2.3$.
They span the full range in $I-K_s$ color. All the galaxies show a break
between the $J_s$ and $K_s$ bands.
The curves show  stellar population fits with either constant
formation and reddening, or unreddened single age bursts.
}
\end{center}}
}

\def\figd{
\vbox{
\begin{center}
\leavevmode
\hbox{%
\epsfxsize=8truecm
\epsffile{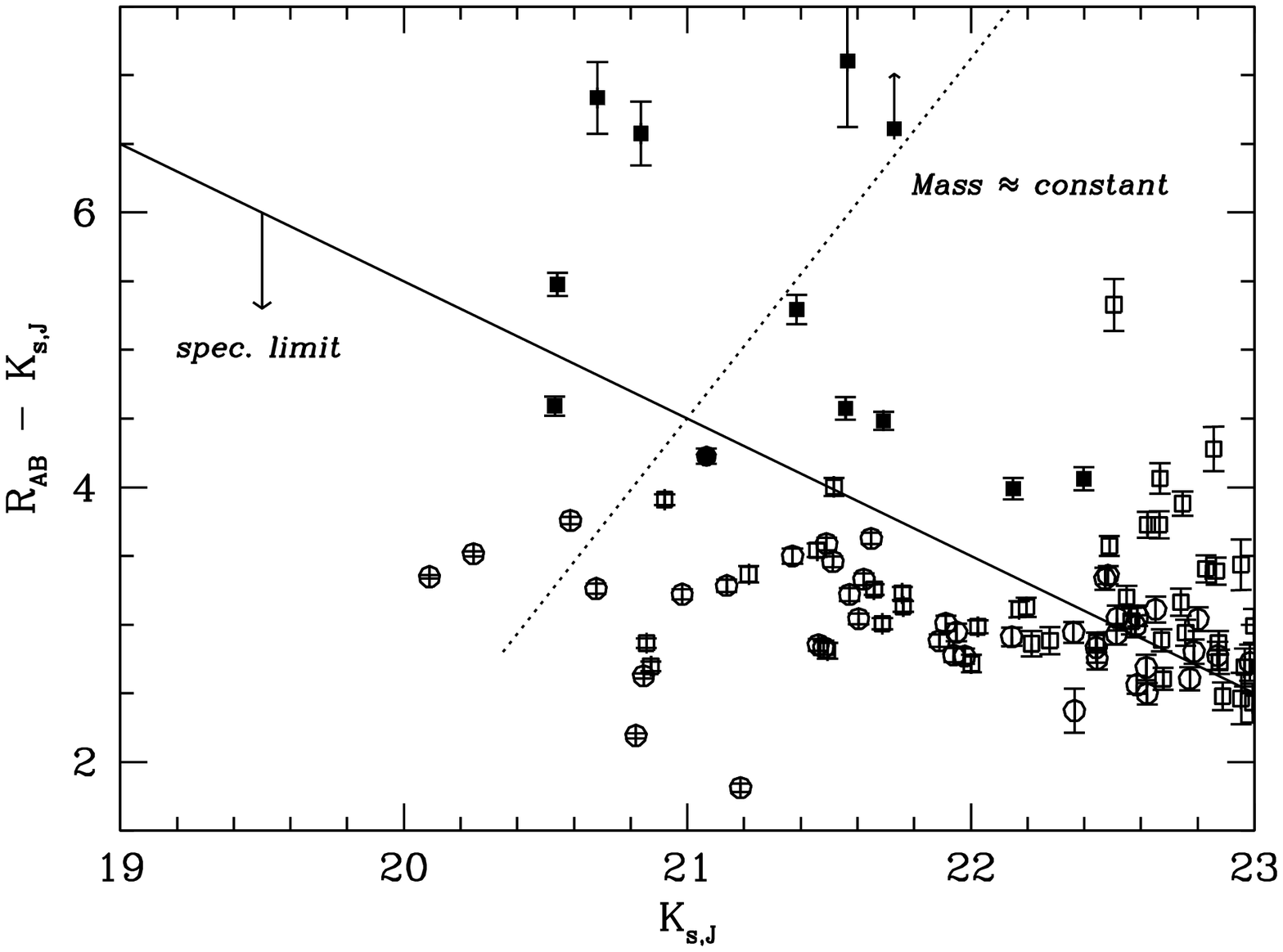}
}
\figcaption{\small
The $R_{AB}-K_s$ color against $K_s$ magnitude for galaxies with $2 <
z < 4$.  The $J_s-K_s$ red galaxies are indicated by filled symbols.
Ly-break galaxies are indicated by circles.  The line indicates the
magnitude limit used for most ground based spectroscopic studies of
Ly-breaks of $R_{AB}=25.5$.  Most $J_s-K_s$ red galaxies are missed
when this magnitude limit is applied.  The dotted line is a track
of constant mass for a model with a
single age population.  As can be seen, the $J_s-K_s$ red galaxies are
slightly less luminous in the $K_s$ band than the brightest Ly-break
galaxies in HDFS, but their masses may well be comparable or higher.
}
\end{center}}
}

One of the most effective ways to find high redshift galaxies is the
Lyman-break selection technique (Steidel \& Hamilton 1993, 
Steidel et al. 1996a,b).
This selection method relies on the strong  break in the
rest-frame far-UV spectrum of strongly star-forming galaxies.
The density of bright Lyman-break galaxies at $z \approx 3$ is comparable
to
that of $L_*$ galaxies nearby, and this population is a major
constituent
of the high redshift universe.
Other selection techniques based on sub-mm, or X-ray emission
have yielded complementary samples, with generally little overlap, and
much smaller number densities
(e.g., Smail et al., 2002a, Barger et al., 2002).
The sub-mm galaxies  may, however,
contribute significantly to the overall star formation rate.

All of these techniques select galaxies which are very
different from normal galaxies in the present-day universe, the light of which
is dominated by rather evolved stars. Normal galaxies are
too large, and too faint in the far-UV to be selected as  Lyman-break
galaxies at $z=3$ (e.g., Giavalisco et al. 1996). 
This raises the question of whether we may still be
missing a significant population of galaxies at high redshift.

This issue can be addressed by means of very deep near-infrared (NIR) imaging.
A normal $L_*$ galaxy would have a K band magnitude of $\approx$23 when
redshifted to $z=3$, out of
reach of 4m class telescopes.
The development of large telescopes with good image quality,
and large NIR detectors has made it possible to achieve
the depth necessary to select high redshift galaxies by their
observed infrared light.

In this paper we present results obtained at the VLT on the Hubble
Deep Field South (HDF-S, Williams et al. 2000, Casertano et al. 2000),
as part of the FIRES project (Franx et al. 2000).
We imaged the WFPC2 field of the HDF-S with ISAAC 
in 3 passbands ($J_s$, $H$, $K_s$), with a total integration time of
103 hours. The data processing and results are discussed in
\Labbe\ et al. (2003). The first results based on a subset of the data
were presented by Rudnick et al. (2001). All magnitudes given here
are in the Johnson system unless noted otherwise.

\section{Infrared  selection of high redshift galaxies}

We consider the selection of galaxies which are not dominated
by an unobscured starburst in their rest-frame UV-optical light.
We wish to define a simple photometric criterion which can select
such galaxies, similar to the U-dropout criterion to select Ly-break galaxies.
Unfortunately, the rest-frame optical spectra of optically red galaxies
do not contain any feature as prominent as the Lyman break of
star-forming galaxies.  However, we can use the Balmer break and 4000
\AA\ break to select high redshift galaxies. The lower strength of
this feature implies that we need to obtain very deep
photometry. Since the breaks shift into the $J_s$ band at 
$z\approx$ 2, a red $J_s - K_s$ color is a 
\figa
\vskip 1pt {\parindent = 0pt
simple and effective
criterion.  This is illustrated in Fig. 1, where we show tracks of
$J_s - K_s$ color versus redshift for model spectra taken from Bruzual
\& Charlot (1996). If we impose a criterion $J_s - K_s > 2.3$, we
select galaxies with $z > 2$, even if we allow for dust extinction.
}

Exceptions will occur, obviously. Some of the lower redshift EROs
have $J-K > 2.3$  (e.g., Smail et al. 2002b), but
their surface density is low ( $\approx$ 0.2
arcmin$^{-2}$). Hall et al. (2001) selected 4 galaxies with $J - K >
2.5$, and found photometric redshifts $z \ge 2$, in agreement with our
simple tracks.

Figure 2 shows the $J_s-K_s$ versus $I-K_s$ colors of the galaxies in 
the HDF-S with $K_s < 22.5$.
We find  14
galaxies down to this limit with  $J_s-K_s > 2.3$.
Since our effective area is 4.7 arcmin$^{2}$, this implies a surface 
density  of 3 $\pm$ 0.8 galaxies arcmin$^{-2}$,
which is seven times that
of bright Lyman-break galaxies around $z=3$ selected
by the ground-based U-dropout technique (Steidel et al. 1996a). 
The redshift distributions of the two samples are significantly
different, however, and the volume densities are more comparable as
we shall see below.


The full spectral energy distributions of a subset of the galaxies are
shown in Fig. 3. The majority of the galaxies show a distinct break
in the NIR, which we interpret as the  Balmer/4000 \AA\
break. The galaxies have a wide range in optical-NIR colors
(Figs. 2 and 3). We find 5  extremely red galaxies with
$J_s - K_s > 3$, and 4 of these are red in all colors. Their colors
are rather similar to the galaxies found by Dickinson et al. (2000),
and Im et al. (2002).

\section{ Redshifts and density}

Most of the galaxies are  extremely faint in the optical, with a
median V band magnitude of  26.6.
As a result, it is very difficult to measure 
redshifts spectroscopically with current instrumentation, and
we have to  rely on photometric redshifts 
\figb\vskip 1pt {\parindent=0pt
instead.
Fortunately, the spectral breaks which many of the galaxies show 
in the NIR help to derive the photometric redshifts.
}

We have used the photometric redshifts published by \Labbe\ et al
(2003), based on algorithm developed by Rudnick et al.
(2001, 2003). The SED is fit with a linear combination of
templates based on observed spectra of nearby galaxies.
It gives good agreement with the spectroscopic redshifts available in 
the HDF-South,
with an average $| \Delta z| / (1+z)  = 0.08$.
We verified that the derived density remains the same within the errors
if another method is used (``hyperz'' by Bolzonella, Miralles,
\& Pello  2000). Spectroscopic confirmation of the photometric
redshifts is urgently needed, obviously.

The photometric redshifts of the red sample with $K_s < 22.5$ 
range from 1.92 to 4.26, with a median of 2.6, and an rms of 0.7.
The  volume density is derived in the interval $2 < z < 3.5$, identical
to that of U-dropouts selected from WFPC2 imaging.
Eleven galaxies lie in this redshift range, resulting in a volume density of 
0.0014 $\pm$ 0.0004 \hmpc3.
Errors in the photometric redshifts are likely to
reduce the measured volume density, as they will push galaxies
outside the sampled volume.
The derived volume density is  half that
of Lyman-break galaxies at $z\approx 3$ to $R_{AB}=25.5$ (Steidel et al. 1999).
Hence the contribution of this sample to the overall density is
substantial.
The variations due to large scale structure are the main source of
uncertainty. We note, for example, 
that the number of galaxies
with red $I - H$  color is much higher in HDF-South than in
HDF-North  (\Labbe\ et al, 2003). On the other hand, our second deep field
of 5x5 arcmin
contains a similar surface density of $J_s-K_s$ selected galaxies as 
HDF-South to
$K_s=21$ (van Dokkum et al, 2003).
Surveys over larger areas are needed to establish how
typical these areas are, and are now in progress.

\section{Comparison with Lyman-break selected samples}

The large volume density derived above raises the question how the
red galaxies are related to the ``classical'' 
Lyman-break 
\figc \vskip 1pt {\parindent=0pt
galaxies (Steidel et al. 1996a,b). In principle, a
galaxy can be both red in $J_s - K_s$ and be a Lyman-break.
In order to compare the two samples we select
galaxies in the HDF-South with
$ 2 < z < 4 $ and $K_s < 22.5$, and apply the criteria for U-dropouts
as defined by Steidel et al. (1996b).
Surprisingly, the two samples are nearly disjoint.
Only 1 galaxy satisfies both criteria.
Many of the $J_s - K_s$ galaxies fail to satisfy the U-drop criteria because
they are too faint in the $B$  or  $V$ band. Even with
HDF-South quality data, it is not possible to establish whether they
have a prominent break between $U$ and $B$.
Conversely, the median $J_s - K_s$ color of the U-dropout
galaxies is very blue: we find a median  of 1.6, identical
to that found by Shapley et al. (2001) for a ground-based selected sample.
}

The disjoint nature of the two populations is illustrated in Fig. 4.,
where we show the $R_{AB} - K_s $ color versus $K_s$ magnitude. The
synthetic $R_{AB}$ magnitude was derived by averaging the $V_{AB}$ and
$I_{AB}$ magnitudes.
Galaxies of different types are indicated by different
symbols. The U-dropout galaxies are blue and follow a well defined
color-magnitude relation, as found earlier by Papovich et al. (2001).
The $J_s - K_s$ selected galaxies show a very large range in color,
$ 4 \le R_{AB} - K_s \le 7$.

The relative contributions of the two types of galaxies to 
the total observed $K_s$ band light is  similar to their relative
number densities:
The galaxies with $J_s - K_s > 2.3$ contribute 24\% to the total 
observed $K_s$ band light from galaxies with
$2 < z < 4$ and $K_s<22.5$,  and the Lyman-breaks
contribute 55 \%.
The relative contribution of the $J_s - K_s$ selected galaxies to the stellar
mass is expected to be higher,
as the red colors indicate higher mass-to-light ratios.
Fig 4. shows the track of constant stellar mass for a single stellar 
population  
observed at $z=3.0$ for ages between 0.1  and 2 Gyr, based on models by
Bruzual \& Charlot (1996).
If we use this track to correct the relative luminosities to relative masses,
we derive that the $J_s - K_s$ selected galaxies contribute
43\% to the stellar mass.
This fraction increases
further if the sample were mass selected.
Hence,  the population of  red galaxies is  a significant
component  at the high mass end.

Fig. 4.  can also be used to analyze the selection effects of
most of the ground-based spectroscopic surveys.
Usually, 
\figd
\vskip 1pt {\parindent = 0pt
spectroscopic samples of U-dropout galaxies are
selected  to have $R_{AB} < 25.5$,  indicated 
by the continuous line.
All but one of the $J_s - K_s$
galaxies have $R_{AB} > 25.5$, and hence most would be excluded.
The $R$ band selection criterion is effective
in selecting galaxies with strong, unobscured  star formation, but
misses a significant population of galaxies with redder colors.
Lines of constant mass run nearly perpendicular to the $R$ band
selection criterion: if we select galaxies in the optical we miss
red galaxies over a wide range of masses, and up to the highest masses.
Follow-up studies of NIR selected samples are necessary to obtain
a full census of the high redshift universe, unfortunately,
such studies are very hard.
}

\section{ Stellar populations}

The  $J_s - K_s$ selected galaxies have much redder colors than most
Lyman-breaks, which  can result from
older ages, more reddening, or a combination of effects.  
Furthermore, emission lines can contribute to some of the passbands
and influence the colors.
Here we attempt to understand the stellar populations of the galaxies
better by analyzing the colors and SED's. For the moment, we ignore
the effects of emission lines as they are likely small
(e.g., Shapley et al. 2001, van Dokkum et al. 2003, van Dokkum et al,
in preparation).

The breaks in the SED's of many of the galaxies can be used to
put lower limits on the ages. 
Color-color diagrams using the $I-J_s$, $J_s - H$, and
$H-K_s$ colors show that 9 out of 14 galaxies have breaks which cannot
be explained by reddened models of stellar populations with constant
star formation and young age ( $\le$100 Myr). Generally, models with
ages  $>$ 300 Myr and higher can produce strong enough breaks to explain
the features, in combination with dust. 

We can use the full SED to fit models of stellar populations. We used
models by Bruzual \& Charlot (1996), with a Calzetti et al. (2000) 
reddening law.
As is well
known, it is very hard to disentangle the effects of age and
dust in such fits (e.g., Papovich et al. 2001, Shapley et al. 2001).
However, 11 out of 14 of the galaxies are badly fit by
models with very young populations ( age $\le$100 Myr) and very large
reddening. The $\chi^2$-test rules out these fits at the 95\%
confidence level. Unreddened single age bursts can be ruled out in 
the same way for 12 galaxies.
Models with constant
star formation produce a median age of 1 Gyr and a
median reddening of E(B-V)= 0.5. Both the age and the reddening are
about 3 times higher than the median values
for Lyman-breaks at $z=3$, 0.3 Gyr and 0.16, respectively
(Shapley  et al. 2001).
Two of
the very red galaxies are best fit by single burst models 
without dust, as determined from the $\chi^2$.
These models have an age of 3.5 Gyr.
If all galaxies are fit with a single age population,
the median age derived is 0.7 Gyr, with a large spread.  

As shown by Papovich et al (2001) and Shapley et al (2001), the
derived ages and reddening can vary by a factor of 2 if other
star formation histories, IMF's, or reddening laws are used.
The relative ages of galaxies are more stable, however.
Our models are identical to those used by
Shapley et al (2001), hence the age comparison should be more secure.
Again, follow-up studies are required to constrain the stellar 
populations better. NIR spectroscopy will be essential to estimate
star formation rates and reddening from emission lines.

\section{Discussion}

The results presented here raise many questions. Obviously, follow-up work is
required to verify how typical the HDF-South is. 
Deep NIR-imaging over a wide area is needed to obtain a robust estimate of
the surface density.
Furthermore, it is critical to 
determine the spectroscopic
redshifts of at least some of these sources. The first spectroscopic
confirmation of galaxies selected in this way is presented in a 
companion paper (van Dokkum et al., 2003).
The most pressing question of all, however, concerns the
nature, formation and evolution of these red galaxies, and their relation
to other high redshift galaxies, especially to the
well-studied Lyman-breaks.

The simplest explanation would be that the red galaxies are 
directly related to Lyman-break galaxies. They may be viewed along
dusty lines of sight, or during intermittent epochs of low star formation.
If such a direct relation would hold, one would expect that both
samples have a similar distribution on the sky, similar sizes, and
similar clustering. The first results by Daddi et al. (2003) 
imply  that the clustering of high redshift galaxies increases
rapidly with redder  $J_s - K_s$ colors. If confirmed by spectroscopy
and larger area surveys, this would be  inconsistent with
this very simple scenario.

Alternatively,  the red galaxies could be  a  population  distinct from
the $z \approx 3 $ Lyman-breaks.
They may have been Lyman-breaks at higher redshifts,
and have become redder because of a decline in the star formation, an increase
in age, and an increase of metallicity, and thereby dust. All three
factors are expected to play a role in realistic galaxy evolution
models.
In this context, it is interesting to note that models of observed Lyman-break
galaxies
indicate that such red descendants must exist, as the ages of the
stellar
populations are generally much younger than the age of the universe
at the redshift of the galaxies
(e.g., Papovich et al., 2001, Ferguson et al. 2002).
Unless these galaxies evolve in unexpected ways, one must see older, and
relatively red ``remnants''. 

In the nearby universe, the reddest galaxies are generally the most
massive, with the highest ages, metallicity, and correlation length.
It is possible that such a relation already existed in the early
universe.
The red galaxies found here would be the descendants of galaxies which
started to form  stars very early, and would be
related to more massive halos than those of the young Lyman-break
galaxies at $z=3$.
In this case their spatial densities are expected to be higher, and 
their correlation
length to be enhanced, as found by Daddi et al. (2003).
Work is in progress to test these predictions further.
Unfortunately, the physical processes which determine the appearance of
high redshift galaxies are not well understood, e.g., why are many
strongly  star-forming galaxies
at $z=3$ UV-bright, and most counterparts at low redshift strongly
reddened ? In the absence of a good understanding of the high redshift
galaxies, and
of the feedback processes which may regulate star formation, it is
not clear whether a simple extrapolation of the phenomena at low
redshift
to high redshift is valid.

Finally, caution is required, and we should leave open the possibility
that we do not understand the nature of these sources yet.
We note that similar objects have been found by others (e.g.,
Dickinson et al. 2000, Hall et al. 2001, Totani et al. 2001,
Im et al. 2002). 
Although we have obtained
spectroscopy for a limited sample (van Dokkum et al., 2003), this
is only  feasible for a small fraction of the sample at this moment.
Near-infrared spectroscopy is very urgently needed to confirm that the 
colors which
are observed are not dominated by emission lines, although it appears
quite unlikely that many galaxies would be affected greatly.

\begin{acknowledgements}
The comments of the referee helped to improve the paper.
We thank the staff at ESO for their hard work to take these data and
make
them available. This research was supported by grants from the
Netherlands Foundation for Research (NWO), the Leids Kerkhoven-Bosscha
Fonds, a NASA SIRTF fellowship, and the Lorentz Center.

\end{acknowledgements}

\end{document}